\documentclass[twocolumn,showpacs,amsmath,amssymb,prl]{revtex4}

\usepackage{graphicx}
\usepackage{bm}

\begin{document}

\title{Spectrum of bound fermion states on vortices in $^3$He-B.}

\author{M.\,A.\,Silaev}

\affiliation{ Institute for Physics of Microstructures RAS, 603950
Nizhny Novgorod, Russia.\\  Low Temperature Laboratory, Helsinki
University of Technology, 02150 Espoo, Finland.}

\date{\today}

\begin{abstract}
 We study subgap spectra of fermions localized within vortex cores in $^3$He-B.
 We develop an analytical treatment of the low-energy states and
 consider the characteristic properties of fermion spectra for different types of vortices.
 Due to the removed spin degeneracy the spectra of all singly quantized
 vortices consist of two different anomalous branches crossing the Fermi level. For singular $o$ and $u$ vortices
 the anomalous branches are similar to the standard Caroli-de Gennes -Matricon ones and intersect
 the Fermi level at zero angular momentum yet with
 different slopes corresponding to different spin states.
 On the contrary the spectral branches of nonsingular vortices intersect the Fermi
 level at finite angular momenta which leads to the appearance of a large number of
 zero modes, i.e. energy states at the Fermi level. Considering the $v$, $w$ and $uvw$ vortices
   with superfluid cores we show that the number of zero modes is
   proportional to the size of the vortex core.
\end{abstract}

 \maketitle



 {\bf 1. Introduction.}
 Since the pioneering work of Caroli, de Gennes and Matricon (CdGM) \cite{CdGM} it is well known that
 quantized vortices in superconductors and Fermi suprfluids have a
 non trivial internal electronic structure. It consists of low energy fermionic excitations
 localized within the vortex cores with characteristic interlevel
 spacing defined as $\Delta_0^2/E_F\ll\Delta_0$, where
 $\Delta_0$ is the energy gap far from the vortex line and $E_F$ is the Fermi energy.
 For conventional s-wave superconductors the excitation spectrum of each
 individual vortex $E(Q)$ of a subgap state varies from $-\Delta_0$ to
$+\Delta_0$ as  one changes the angular momentum $Q$ defined with
respect to the vortex axis.

 At small energies
$|E|\ll\Delta_0$ the spectrum is a linear function of $Q$:
 \begin{equation}\label{CdGMspectrum}
  E(Q)\simeq-Q\omega,
 \end{equation}
 where $\omega\approx\Delta_0/(k_\perp\xi)$,
$\Delta_0$ is the superconducting gap value far from the vortex
axis, $k_\perp=\sqrt{k_F^2-k_z^2}$, $k_F$ is the Fermi momentum,
$k_z$ is the momentum projection on the vortex axis,
 $\xi=\hbar V_F/\Delta_0$ is the coherence length, $V_F$ is
the Fermi velocity, and $Q$ is half an odd integer.  Under some
exotic conditions \cite{Geim} several vortices can merge and then
one obtains a multiquantum vortex with a certain winding number
$M$. The number of anomalous branches per spin
projection~\cite{Volovik-1993} is equal to the vorticity $M$. For
the states with an even vorticity all the anomalous branches cross
the Fermi level at nonzero angular momentum $Q_j$:
\begin{equation}
\label{Volovik-spectr}
  E(Q)\sim-(Q\pm Q_{j})\Delta_0/(k_\perp\xi)\,
\end{equation}
where $j=1 ... M/2$, $Q_{M/2}\sim k_\perp\xi$. For a vortex with
an odd winding number there appears a branch crossing the Fermi
level at zero impact parameter.

The quantized vortices in $^3$He-B have much in common with
vortices in ordinary s-wave superconductors. However in
multi-component superfluid system  $^3$He axial symmetry allows
the nucleation of additional order parameter components inside
vortex core. Thus vortices in this system are in general
nonsingular, i.e. may have a superfluid core unlike singular
vortices in s-wave superconductors which always have a normal
core.  There exist five types of vortices with different internal
core structures in $^3$He-B: $o$, $u$, $v$, $w$ and $uvw$ vortices
\cite{VolovikHe3BPrl,VolovikHe3B,VolovikRMP}. The $o$ vortex is
the most symmetric one, it has no superfluid core and consists of
almost pure B-phase without inclusions of other phases. Other
vortices break some of the discrete symmetries existing for the
most symmetric $o$ vortex. Among them the $u$ vortex is singular
while the remain $v$, $w$ and $uvw$ vortices have superfluid
cores.

   According to the analysis in the framework of
Ginzburg-Landau theory \cite{VolovikHe3B,VolovikRMP,Passvogel}
near the critical temperature only $v$ vortex is stable. The cores
of such vortices are occupied by an A phase and a ferromagnetic
$\beta$ phase \cite{VolovikHe3BPrl,VolovikHe3B,VolovikRMP}. These
additional phases correspond to a nonzero total angular momentum
projection on the vortex axis and a zero vorticity in the real
space. Therefore nuclea of additional phases remain finite at the
vortex center. Nucleation of ferromagnetic $\beta$ phase inside
vortex cores explains a large spontaneous magnetic moment of
vortices revealed in the NMR experiments in rotating $^3$He-B
\cite{NMR1}. The first order phase transition seen in the NMR
experiments was associated with the change of the symmetry of the
internal core structure \cite{NMR1,NMR2}.

 As was shown by Volovik
\cite{VolovikGaplessFermions,VolovikLocalizedFermions} in vortices
with dissolved core singularity the spectrum of bound fermion
states can be substantially modified in contrast to ordinary CdGM
spectrum of singular vortices. In particular the presence of other
superfluid phases inside vortex core leads to the appearance of
large number of zero modes, i.e. the spectral branches crossing
the Fermi level. The number of these zero modes can be as high as
$E_F/\Delta_0\sim k_F\xi \gg 1$. Thus a minigap in spectrum of
bound fermions, which is a characteristic feature of CdGM spectrum
\cite{CdGM} is absent for nonsingular vortices in $^3$He-B. Zero
modes also exist even for a singular and most symmetric $o$
vortex. Although the number of them is much smaller than for
nonsingular vortices but it can be effectively controlled by
external magnetic field \cite{VolovikMisirpashaev,
VolovikMakhlin}.  Even in zero magnetic field due to a broken
relative spin-orbital symmetry in B-phase of $^3$He the spin
degeneration of the energy spectrum is removed
\cite{VolovikMisirpashaev}. As a result
 the CdGM spectral branches acquire spin dependent shift which
closes the minigap. Localized fermions which occupy the negative
energy states on the spectral branches intersecting zero energy
level form a one dimensional Fermi liquid inside vortex core,
which can lead to the instability of the vortex core structure
\cite{VolovikMakhlin}.

In this Letter we develop a generalization of CdGM theory for the
case of vortices in B phase of superfluid $^3$He. We derive a
general expression for spectrum of vortex core quasiparticles in
the presence of multiple order parameter components inside vortex
core and analyze the spectra of several particular vortex types.

 The method that we use is based on the approximate analytical solution
of quasiclassical Andreev equation describing the motion of
quasiparticles along the trajectories inside vortex core. Earlier
 this method was applied to study the spectrum of quasiparticles localized within the cores
 of multiquanta vortices \cite{Volovik-1993,MRS-2008}.
 Generally the Andreev equation for two component wave function $\psi=(U,V)$
 along the quasiclassical trajectory
  has the form
\begin{equation}
\label{Quasiclass}
  -i\tilde{\xi}\hat\tau_3\frac{\partial \psi}{\partial s}
  +\hat\tau_1 \hat \Delta_R \psi-\hat\tau_2 \hat\Delta_{Im} \psi
  =E \psi \ ,
\end{equation}
where $\hat\tau_{1,2,3}$ are Pauli matrices of Bogolubov-Nambu
spin, $\tilde{\xi}$ is a length scale of the order of coherence
length, $\hat\Delta_R=(\hat\Delta+\hat\Delta^+)/2$ and
$i\hat\Delta_{Im}=(\hat\Delta^+ - \hat\Delta)/2$ are the hermitian
and anti-hermitian parts of normalized gap operator, $E$ is a
normalized energy with the normalization energy of the order of
bulk value of gap function $\Delta_0$.

Here we should take into account the spinor structure of
quasiparticle wave functions which is essential in $^3$He. In this
case the coefficients $\hat\Delta_{R,Im}$ in Andreev equation
(\ref{Quasiclass}) are $2\times 2$ matrices in spinor space. Then
the matrix equation (\ref{Quasiclass}) is a system of 4 scalar
equations. If matrices $\hat \Delta_{R}$ and $\hat \Delta_{Im}$
commute $[\hat \Delta_{R},\hat \Delta_{Im}]=0$ the fourth order
Andreev equation can be reduced to 2 equations of the second
order. However this can not always be the case. To develop a
general perturbation theory we note we note that if
$\hat\Delta_{Im}\equiv 0$ the exact solution of the Andreev
equation (\ref{Quasiclass}) corresponding to $E=0$ can be obtained
in a spinor basis diagonalizing the matrix $\hat \Delta_{R}=
diag(\Delta_{R1},\Delta_{R2})$. Then we obtain two degenerate
solutions $\psi_{1,2}=(1,-i)_\tau f_{1,2}(s)$ corresponding to the
zero energy, where
$f_{j}(s)=\Lambda_j\exp\left(-\tilde{\xi}^{-1}\int_0^s \Delta_{Ri}
ds\right)$ and $\Lambda_j$ is an eigen spinor of matrix $\hat
\Delta_{R}$. As we will see below in case of a single--quantum
vortex the functions $\Delta_{R1,2}(s)$ have asymptotics of
different signs $\Delta_{Ri}(+\infty)\Delta_{Ri}(-\infty)<0$. We
assume that $\Delta_{Ri}(+\infty)>0$ therefore the solutions
$f_{1,2}(s)$ decay at $s=\pm\infty$. Using this localized solution
as a zero--order approximation for the wave function the spectrum
can be found within the first order perturbation theory assuming
that $|E|\ll 1$ and $|\hat\Delta_{Im}(s)|\ll 1$. In general
$f_{1,2}(s)$ are not the eigen spinors of the operator $\hat
\Delta_{Im}$ which therefore couples the $\psi_1$ and $\psi_2$
states. Then the standard perturbation theory yields the secular
equation
 \begin{equation}\label{Spectrum-alpha-generalized}
  \det \begin{pmatrix}
    S_{11}-E & S_{12} \\
    S^*_{12} & S_{22}-E \\
  \end{pmatrix}=0
 \end{equation}
 where the matrix elements are $S_{11(22)}=2\langle f_{1(2)}|\hat \Delta_{Im}|f_{1(2)}\rangle$
 and $S_{12}=2\langle f_{1}|\hat \Delta_{Im}|f_{2}\rangle$.

 In general the accuracy of the first order perturbation correction
should be determined by the factor $O(\hat\Delta_{Im}^2)$, where
$|\hat\Delta_{Im}|\ll 1$ is a small parameter. However in a
particular case of Eq.(\ref{Quasiclass}) the second order
correction to the zero energy level is exactly zero and therefore
the accuracy of Eq.(\ref{Spectrum-alpha-generalized}) is much
better: $O(\hat\Delta_{Im}^3)$. To prove this result we assume for
simplicity that $[\hat \Delta_{R},\hat \Delta_{Im}]=0$ so that
$S_{12}=0$. Then if the eigen function $\psi=(U,V)_\tau$ of
Eq.(\ref{Quasiclass}) with $\hat\Delta_{Im}=0$ corresponds to the
energy $\varepsilon_n$ the other function
$\tilde{\psi}=(-V,U)_\tau$ corresponds to the energy
$-\varepsilon_n$. Therefore it is easy to check that the
contribution from negative energy levels to the second order
perturbation of the energy level $E=0$ exactly compensates the
contribution from the positive levels. The proof modification to
the general case $[\hat \Delta_{R},\hat \Delta_{Im}]\neq 0$ is
straightforward.

 {\bf 2. Basic formulas}.
Our further consideration is based on the Bogoulubov- Nambu
equation for the quasiparticles near the Fermi level. From the
beginning we assume the system to be homogeneous in $z$ direction
which coincides with the vortex axis. Then we obtain
two-dimensional Bogoulubov- Nambu equations  with the effective
Fermi energy $E_\perp=E_F-\hbar^2k_{z}^2/2m$ and the Fermi
momentum in $xy$ plane $k_\perp=\sqrt{k_F^2-k_z^2}$:
\begin{equation}\label{BN}
\hat H_0\psi +\hat\tau_1  \hat\Delta_R\psi- \hat\tau_2
\hat\Delta_{Im}\psi =E\psi \ ,
\end{equation}
where $\hat\tau_j$ ($j=1,2,3$) are Pauli matrices in particle-hole
space, $\hat H_0=\hat\tau_3(\hat {\bf p}^2-\hbar^2k_{\perp}^2
)/2m$, and $\hat {\bf p}=-i\hbar\nabla$. Further we will assume
that the gap function and energy are normalized to the bulk value
of the energy gap $\Delta_0$.

Generally the gap function in $^3$He-B can be parameterized as
follows: $ \hat\Delta=-i\hat \sigma_y ({\bf \hat \sigma}\cdot{\bf
d})$, where ${\bf d}$ is a vector in 3D space and $\sigma_{1,2,3}$
are Pauli matrices in conventional spin space. Being
 proportional to the wave function of Cooper pairs in isotropic liquid $^3$He
 the gap function can be presented as a superposition: $
 \hat\Delta=\sum_{\mu,\nu}C_{\mu\nu}\{e^{i(M-\mu-\nu)\varphi},\hat\Delta_{\mu\nu}\}$,
 where $\varphi$ is a polar angle in $xy$ plane,
  $M$ is vorticity, $\hat\Delta_{\mu\nu}$ is a Cooper pair wave function
 with definite angular momentum $\nu=-1,0,1$ and spin $\mu=-1,0,1$ projections on the $z$
 axis and $\{\hat A \hat B\}=(\hat A \hat B+\hat B \hat A)/2$ is an anticommutator.

 The order parameter distribution should be axisymmetric with the
 generator of rotation symmetry around $z$ axis \cite{VolovikRMP}
 $\hat Q= \hat L_z+\hat S_z-M\hat I$, where $\hat L_z$ and $\hat S_z$ are
 the projections of internal angular momentum and spin
 of Cooper pairs onto the $z$ axis and $M$ is a total vorticity.
 Thus for all order parameter components
 the condition $\mu+\nu=M$ should be satisfied.
 For singly quantized vortices $M=1$ there can exist five basic components of the order
 parameter. Among them are $C_{1,-1}$, $C_{-1,1}$ and $C_{00}$
  which correspond to the main B phase, $C_{0,1}=C_{A}$ and
$C_{1,0}=C_{\beta}$ which correspond to the
 additional A and $\beta$ phases localized inside vortex core. The additional A phase
 has a zero spin projection ($\mu=0$) and unit projection of
 orbital momentum ($\nu=1$) on the $z$ axis while $\beta$ phase has $\mu=1$ and $\nu=0$.
 The components of gap function $\hat\Delta_{\mu\nu}$
 are characterized by ${\bf d}$ vector as follows \cite{VolovikRMP}
 $ {\bf d}= {\bf \lambda}^\mu({\bf \lambda}^\nu\cdot {\bf
 q})$, where ${\bf q}={\bf k}/k_F$ and ${\bf \lambda}^{\pm 1}=({\bf x_0}\pm i{\bf
 y_0})$, ${\bf \lambda}^0={\bf z_0}$. Correspondingly in
 B phase we have ${\bf d}={\bf q}$, in A phase ${\bf d}={\bf
 z_0}(q_x+iq_y)$ and in $\beta$ phase ${\bf d}=q_z ({\bf x_0}+i{\bf
 y_0})$. Far from the vortex core at $r\gg\xi_v$ only B superfluid phase
 exists so that $C_{1,-1}=C_{-1,1}=C_{00}=1$
  and $C_{A,\beta}=0$. The vortex type is determined by the behaviour of amplitudes $C_{\mu\nu}$
  at smaller distances $r\sim\xi_v$ and there exist five types of vortices \cite{VolovikRMP}.

   Vortices of $o$ and $u$ types are singular so
  that only the superfluid components of B phase $C_{1,-1}$, $C_{-1,1}$ and $C_{00}$
  are nonzero.
   These amplitudes are real for the most symmetric $o$ vortex and
 complex ones for $u$ vortex with conserved parity $P_1=P$ but
 broken $P_3=TO^{\pi}_x$ discrete symmetry. Here $T$ is time inversion and
  $O^{\pi}_x$ is a rotation
  by the angle $\pi$ around the axis $x$ perpendicular to the
 vortex axis $z$. The gap function which describes singular B
 phase vortices can be presented in the form
  \begin{equation}\label{DeltaB0}
 \hat\Delta_B=-i\hat\sigma_y\{({\bf \hat\sigma}\cdot{\bf
d_B}),e^{i\varphi}\}
 \end{equation}
 where ${\bf d_B}=(B_+q_x-iB_-q_y,B_+q_y+iB_-q_x,C_{00} q_z)$,
 $B_{\pm}=(C_{1,-1}\pm C_{-1,1})/2$.  Generally
 $B_{\pm}=B_{\pm}(r)$ are arbitrary complex functions
 with asymptotics
 $B_{+}(\infty)=1$ and $B_-(\infty)=0$ so that ${\bf d_B}(\infty)={\bf
 q}$.

Nonsingular $v$, $w$ and $uvw$ vortices have superfluid cores with
the inclusion of A and $\beta$ phases:
 \begin{equation}\label{DeltaA0}
 \hat\Delta_{A}=C_{A}q_\perp e^{i\theta_p}\hat\sigma_x,
 \end{equation}
 \begin{equation}\label{Delta-beta0}
 \hat\Delta_{\beta}=C_{\beta}q_z\left(1-\hat\sigma_z\right),
 \end{equation}
 where $q_\perp=k_\perp/k_F$.
 The functions $C_{A,\beta}=C_{A,\beta}(r)$ describing the spatial distributions of additional A and $\beta$
  phases inside vortex core are finite at $r=0$ and vanish outside the core at $r\gg\xi_v$.
 The $v$ and $w$ vortices are characterized by real B phase
 amplitudes. If $C_{A,\beta}$ are also real
  then we have a $v$ vortex with conserved $P_2=PTO^{\pi}_x$ symmetry. The case when $Re(C_{A,\beta})=0$, $Im(C_{A,\beta})\neq 0$ corresponds
 to $w$ vortex with conserved $P_3=TO^{\pi}_x$ symmetry. The less
 symmetric $uvw$ vortex with all discrete symmetries $P_1,P_2,P_3$ broken
 has complex amplitudes of B, A and $\beta$ phases.

Within the quasiclassical approximation Eq.(\ref{BN}) can be
reduced to 4 equations of the first order along
 linear trajectories, i.e. the straight lines along the direction
 of Fermi momentum
 ${\bf{q}}=(\cos\theta_p,\sin\theta_p)$
 (for a detailed review of this transformation see e.g.
Ref.\cite{MRS-2008}). Each
 trajectory is specified
  by the angle $\theta_p$ and the impact parameter
 $b=k_F{\bf z_0}\cdot ({\bf q}\times {\bf r})$. Introducing the coordinate
 along trajectory $s=({\bf{q}}\cdot{\bf{r}})$ we
 arrive at the quasiclassical equation $\hat H\psi =E \psi$
 for the  wave function $\psi(s,\theta_p)$. The quasiclassical hamiltonian is
\begin{equation}
\label{bdg} \hat H = -i\xi q_\perp
\hat\tau_3\frac{\partial}{\partial s}+ \hat\tau_1 \hat\Delta_R-
\hat\tau_2  \hat\Delta_{Im}.
\end{equation}

The impact parameter of quasiclassical trajectories is
proportional to the projection of angular momentum $Q$ of
quasiparticles on the $z$ axis: $b=-Q/k_\perp$. However the
hamiltonian (\ref{bdg}) does not commute with the corresponding
operator $\hat L_z=-i\partial/\partial \theta_p$ since in general
it is not conserved in $^3$He. Still due to the axial symmetry of
vortices the total momentum $\hat L_z+\hat S_z$ is conserved.
Therefore the angular and coordinate variables $\theta_p$ and $s$
in the quasiclassical hamiltonian (\ref{bdg}) can be separated.
Let us introduce the new functions
 \begin{equation}\label{U-transform}
  \tilde{U}=e^{i\hat\sigma_z\theta_p/2}\hat M_0^+ U,
 \end{equation}
 \begin{equation}\label{V-transform}
 \tilde{V}=e^{i\hat\sigma_z\theta_p/2}V,
 \end{equation}
 where $\hat M_0=-ie^{i\theta_p}\hat \sigma_y({\bf \hat\sigma}\cdot {\bf
 q})$.
 It is easy to check that the resulting gap operators (\ref{DeltaB},\ref{DeltaA},\ref{Delta-beta}) after this
 transformation  do not depend on the angle $\theta_p$.
For the singular part of gap function we obtain
 \begin{equation}\label{DeltaB}
  \hat\Delta_B=\frac{\left(s-i b\right)}{r}\left[
  C_B+\delta\hat\Delta_{B1}\right],
 \end{equation}
 where $r=\sqrt{s^2+b^2}$, $C_B=q_\perp^2B_+ +q_z^2C_{00} $ and
 \begin{equation}\label{dDelta1}
 \delta\hat\Delta_{B1}=
 q_\perp\left[B_-\left(q_z\hat\sigma_x-q_\perp\hat\sigma_z\right)+
 i(B_+-C_{00})  q_z\hat\sigma_y  \right].
 \end{equation}
 The expressions for the gap function describing the additional A
 and $\beta$ phases inside vortex core read
 \begin{equation}\label{DeltaA}
  \hat\Delta_A = C_A \left( q_zq_\perp-i
  q^2_\perp\hat\sigma_y\right).
 \end{equation}
 and
\begin{equation}\label{Delta-beta}
  \hat\Delta_\beta = C_\beta q_z\left(1-\hat\sigma_z\right)
  \left(q_\perp+ q_z\hat\sigma_x\right).
 \end{equation}
 Then one can search the solution in the factorized
 form: $\psi(s,\theta_p)=\psi(s)\exp(i(Q+1/2)\theta_p)$. Note that from
 Eqs.(\ref{U-transform},\ref{V-transform}) it follows that the values
 of azimuthal quantum number $Q$ should be integer to pertain the unambiguity of initial
 wave function $\psi=(U,V)$.

 {\bf 3. Spectrum of vortex core states.}
 At first let us consider the quasiparticle spectrum of singular
 $o$ and $u$ vortices when $C_{A,\beta}=0$.
 In general case $C_{1,-1}\neq C_{-1,1}\neq C_{0,0}$ the gap
function is given by Eqs.(\ref{DeltaB},\ref{dDelta1}). We will
assume the simplifying condition to be fulfilled  $|B_-|,
|B_+-C_{00}|\ll 1$ which is justified by Ginzburg-Landau
calculations \cite{VolovikRMP}. Then we can take into account only
the hermitian part of the operator
$(\delta\hat\Delta_{B1})_R=(\delta\hat\Delta_{B1}+\delta\hat\Delta_{B1}^+)/2$
in Eq.(\ref{DeltaB}) since the anti hermitian part gives the
contribution to the energy spectrum of the higher order in small
parameter $O(|B_-|,|B_+-C_{00}|)$. In this case the zero order
solution of Andreev equation (\ref{Quasiclass}) is spin degenerate
$\psi_{1,2}(s)=\psi_0(s)=(1,-i)_\tau f_0(s)$.

 Diagonalizing the gap function $\hat
 \Delta_B$ by spin and using the Eq.(\ref{Spectrum-alpha-generalized})
 we obtain the energy spectrum in the following form
   \begin{equation}\label{ou-vortex-Spectrum}
  E(Q,q_z,\chi)=-\omega_{\pm} Q,
 \end{equation}
 with $\omega_\pm=Re\langle C_B/(k_\perp r)\rangle_0+
 \chi F$, where $F^2=k_F^{-2}\left[Re^2 \langle B_-/r\rangle_0+q_z^2 Im ^2\langle
 (B_+-C_{00})/r\rangle_0\right]$ and $\chi=\pm
1$ corresponds to different spin states. For brevity we have
denoted $\langle X \rangle_0=2\langle f_0|X|f_0\rangle$.
   For the $o$ vortex with
$Im (B_+-C_{00})=0$ the difference between $\omega_+$ and
$\omega_-$ is determined by the asymmetry of amplitudes $C_{1,-1}$
and $C_{-1,1}$. For the $u$ vortex the condition is less
restrictive since even in case $C_{1,-1}= C_{-1,1}$ but $
C_{-1,1}\neq C_{00}$ we can obtain that $\omega_\pm$ are
different.

 Now we proceed with the analysis of the quasiparticle spectra
for nonsingular vortices. Here we focus on the influence of the
additional order parameter components and therefore assume the
most simple form of singular part of gap function (\ref{DeltaB})
with $C_{1,-1}= C_{-1,1}= C_{00}$ and consequently
$\delta\Delta_{B1}=0$. At first we consider only the influence of
A phase and put $C_\beta=0$. In this case we have
 \begin{equation}\label{Delth}
 (\hat
 \Delta_A)_{R}=Re(C_A)q_zq_\perp+Im(C_A)q_\perp^2\hat\sigma_y,
 \end{equation}
  \begin{equation}\label{DeltaAh}
 (\hat \Delta_A)_{Im}=\left[Im (C_A)
 q_\perp q_z-\hat\sigma_y q_\perp^2Re(C_A)\right].
 \end{equation}
The matrix coefficients $\hat\Delta_{R,Im}$ in Andreev equation
(\ref{Quasiclass}) are diagonalized simultaneously in spinor basis
 $f_{1,2}\sim (1,\pm i)_\sigma$ so that $\hat\sigma_y f_{1,2}=\pm
 f_{1,2}$.
  Then Eq.(\ref{Spectrum-alpha-generalized}) yields then the
 following energy spectrum
 \begin{equation}\label{vw-A-phase-Spectrum}
  E(Q,q_z,\chi)=-\omega_j Q+\alpha_{j} q_\perp q_z+\chi \gamma_j q_\perp^2,
 \end{equation}
 where $j=1,2$ corresponds to $\chi=1,-1$,
 $ \omega_j=2\langle f_j|C_B/(k_\perp r)|f_j\rangle$ and
 $\alpha_j,\gamma_j=2Re, Im \langle f_j| C_A| f_j \rangle$.
For $v$ vortex with $Im(C_A)=0$ and $w$
 vortex with $Re(C_A)=0$ it is easy to check that
 $(\omega,\alpha,\gamma)_1=(\omega,\alpha,\gamma)_2$
 in Eq.(\ref{vw-A-phase-Spectrum}) which means following symmetry of $v$ vortex spectrum
$E(Q,q_z,\chi)=-E(-Q,q_z,-\chi)$. Thus the spectrum consists of
two symmetrical anomalous branches having in particular equal
slopes as functions of $Q$ at the Fermi level $E=0$. The spectrum
of $w$ vortex is spin divergent since $\gamma _j=0$ in
Eq.(\ref{vw-A-phase-Spectrum}).
 The situation is more complicated
for the $uvw$ vortex with $Re,Im(C_A)\neq
 0$. In this case $(\omega,\alpha,\gamma)_1\neq(\omega,\alpha,\gamma)_2$
 and the two anomalous branches are not symmetrical.
 Nevertheless there is a relation $(\omega,\alpha,\gamma)_1(q_z)=(\omega,\alpha,\gamma)_2(-q_z)$
 which provides spectrum symmetry $E(Q,q_z,\chi)=-E(-Q,-q_z,-\chi)$
 corresponding to the CPT invariance of Bogolubov-Nambu hamiltonian.

\begin{figure}[h!]
 \centerline{\includegraphics[width=1.0\linewidth]{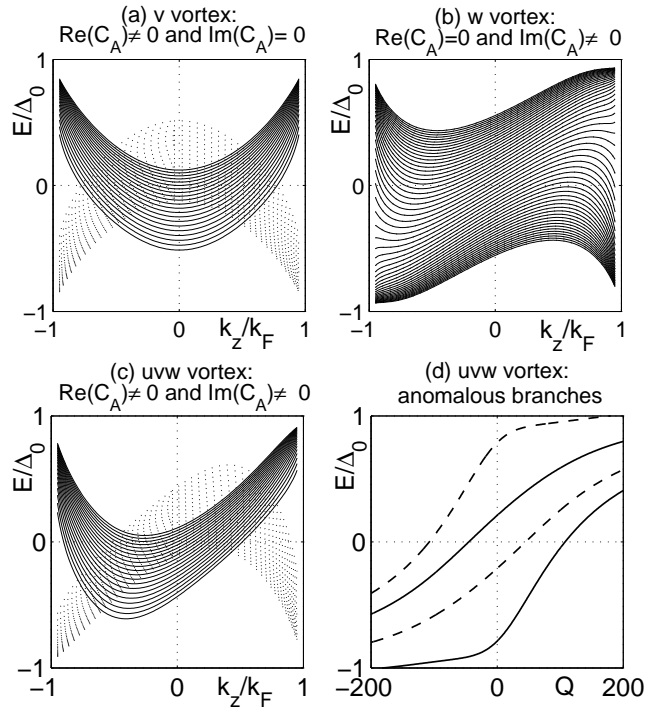}}
 \caption{\label{Spectrum} (a)-(c): Several spectral branches corresponding to
 different spin projections $\chi=1$ (solid lines) and $\chi=-1$ (dash lines).
   (d): Anomalous branches of $uvw$ vortex spectrum. Solid and
   dash lines correspond to the opposite $q_z$ values.  }
   \end{figure}

   In Fig.(\ref{Spectrum}a,b,c) we plot several spectral branches for
 $v$, $w$  and $uvw$ vortices.
We take the model dependencies of the amplitudes of $B$
 and $A$ components in the following form:
 \begin{equation}\label{CBmodel}
  C_B(r)=r/\sqrt{r^2+\xi_{v}^2}
\end{equation}
\begin{equation}\label{CAmodel}
  C_A(r)=e^{i\kappa}\xi_{v}/\sqrt{r^2+\xi_v^2},
 \end{equation}
 where $\kappa$ is an arbitrary phase and $\xi_{v}$ are a characteristic sizes
  so that at $r\gg\xi_{v}$ the asymptotic behaviour is
 $C_B(r)\rightarrow 1$ and  $C_A(r)\rightarrow 0$. In
 Fig.(\ref{Spectrum}d) the two asymmetrical anomalous branches sre
 shown for the $uvw$ vortex. Note that the CPT invariance is retained
 with the help of  anomalous branches corresponding to the opposite
 $q_z$ values which are shown in Fig.(\ref{Spectrum}d) with solid
 and dash lines.

In general case when both $A$ and $\beta$ phases are present the
spectrum of $v$ vortex still consists of two symmetrical branches.
When $C_{A,\beta}$ are real the spin structure of zero order wave
functions is $f_{1,2}\sim(q_z,q_\perp\mp
 1)_\sigma$ so that the hermitian part of gap operator is diagonal in
 this basis:
 \begin{equation}\label{Deltah-v}
 (\hat \Delta_A+\hat
 \Delta_{\beta})_{R}=(C_A+C_\beta)q_zq_\perp+\chi q_zC_\beta,
 \end{equation}
 where $\chi=\pm 1$ corresponds to $f_{1,2}$.
 The anti-hermitian part is
 \begin{equation}\label{DeltaAh-v}
 i(\hat \Delta_A+\hat
 \Delta_{\beta})_{Im}=-i\hat\sigma_y\left(q_\perp^2C_A-q_z^2C_\beta\right).
 \end{equation}
  The spectrum given by Eq.(\ref{Spectrum-alpha-generalized}) differs from
 Eq.(\ref{vw-A-phase-Spectrum}) in the absence of $\beta$ pahse:
 \begin{equation}\label{v-A+beta-spectrum}
  E(Q,q_z,\chi)=-\omega_+
  Q+\chi\sqrt{\omega_-^2Q^2+|S_{12}|^2},
   \end{equation}
 where $\omega_{\pm}=(\omega_1\pm\omega_2)/2$ and
   $S_{12}=2\langle f_1|(\hat \Delta_A+\hat
 \Delta_{\beta})_{Im}|f_2\rangle$. However it is easy to check that
spectral branches are even functions of $q_z$ therefore the
spectrum consists of two symmetrical anomalous branches as before.

 An analogous procedure for the $w$ vortex with
 $Re(C_{A,\beta}=0)$ yields the spectrum consisting of two spin splitted anomalous
 branches
 \begin{equation}\label{w-A+beta-spectrum}
 E(Q,q_z,\chi)=-\omega Q+\gamma +\chi |S_{12}|,
 \end{equation}
 where $\omega=\omega_1=\omega_2$  and $\gamma=2\langle f_1|(\hat \Delta_A+\hat
 \Delta_{\beta})_{Im}|f_1\rangle$. In this case
 $|S_{12}|(q_z)$ is even and $\gamma(q_z)$ is odd function of
 $q_z$. Thus the presence of $\gamma$ phase removes the spin degeneracy of the
 $w$ vortex spectrum. Two spin splitted anomalous branches are not
 symmetrical, since $E(Q,q_z,\chi)\neq -E(-Q,q_z,-\chi)$ similarly
 to the case of $uvw$ vortex.

 The spectra of bound fermion states of nonsingular vortices
 given by Eqs.(\ref{vw-A-phase-Spectrum}, \ref{v-A+beta-spectrum},\ref{w-A+beta-spectrum})
  consist of two anomalous branches crossing the Fermi level at
 some points $Q=Q_{1,2}\neq 0$. Such situation is also realized
 for the spectrum of doubly quantized vortices in ordinary s-wave
 superconductor [see Eq.(\ref{Volovik-spectr})]. An important
 consequence of this fact is an
 existence of zero modes, i.e. spectral branches crossing the Fermi level
 as one can see in Fig. (\ref{Spectrum}).
  Further we discuss zero modes in more detail.

 {\bf 4. Zero modes.}
 As it was shown in Ref.\cite{VolovikLocalizedFermions}
  the number of zero modes $N_0$ strongly depends on the vortex core size $\xi_v$.
  Now with the help of Eq.(\ref{vw-A-phase-Spectrum})
 we will analyze this dependence $N_0(\xi_v)$ for all three types of
 nonsingular vortices in a model situation when the $\beta$ phase
 is absent $C_\beta=0$.
  Let us consider the distributions (\ref{CBmodel},\ref{CAmodel}) of B and A
order parameter components inside vortex core. Such choice of
functions $C_{A,B}(r)$ leads to a simplification of the analysis.
Indeed in this case we have $\alpha_j/\omega_j= q_\perp(k_F\xi_v)
\sin\kappa$ and $\gamma_j/\omega_j=q_\perp(k_F\xi_v)\cos\kappa$.
Then the right hand side (r.h.s.) of the following equation for
zero modes does not depend on $Q$:
\begin{equation}\label{ZM1}
 \frac{Q}{k_F\xi_v}= q_\perp^2\left(q_z\sin\kappa+\chi q_\perp\cos\kappa\right).
 \end{equation}
It is easy to see that for each $Q\neq 0$ the number of
intersections with Fermi level is two or zero. Thus the number of
zero modes is $N_0=4(Q_{min}-Q_{max})$, where $Q_{min}/(k_F\xi_v)$
and $Q_{max}/(k_F\xi_v)$ are minimum and maximum of the r.h.s. of
Eq.(\ref{ZM1}). The additional factor of $2$ is gained from the
summation over the two spin states.
 Then it is obvious that the number of zero modes
 is $N_0= \lambda (k_F\xi_v)$, where $\lambda\sim 1$ is a
constant coefficient.

 {\bf 5. Summary.}
To summarize we have studied the spectra of bound fermion states
localized within vortex cores for different types of vortices in
$^3$He-B. In contrast to vortices in ordinary s-wave
superconductor the spectra of singly quantized vortices in
$^3$He-B in general consist of two anomalous branches
corresponding to the different spin structure of quasiparticle
states. This results from the removing of spin degeneracy of a
standard Caroli-de Gennes - Matricon spectrum. The structure of
two anomalous branches is determined by the vortex type.

The spectrum of singular $o$ and $u$ vortices given by
Eq.(\ref{ou-vortex-Spectrum}) is the most similar
 to the CdGM one (\ref{CdGMspectrum}). However
as distinct from the latter it consists of two anomalous branches
with different slopes. As it follows from Ginzburg-Landau
calculations the asymmetry between pairing amplitudes $C_{1,-1}$,
$C_{-1,1}$ and $C_{00}$ within vortex core is small, therefore the
slope difference should also be small: $|\omega_+-\omega_-|\ll
\omega_{\pm}$.

The spectra of nonsingular $v$, $w$ and $uvw$ vortices consist of
 two spin splitted anomalous branches which intersect the Fermi
level at finite values of angular momenta $Q_{1,2}\neq 0$.  In
case of $v$ vortex the spectral branches are even functions of
$q_z$ within the same spin subband $E(Q,q_z)=E(Q,-q_z)$ which
makes the spectrum analogous to that of the doubly quantized
vortex in s-wave supercondictor [see Fig.(\ref{Spectrum}a)].

 For $w$ and $uvw$ vortices the spectrum can be a general
 function of $q_z$ as it is shown in Figs.(\ref{Spectrum}b,c).
 Note that the "skew" of spectral branches of $w$ and $uvw$ vortices
  is produced by the second term in Eq.(\ref{vw-A-phase-Spectrum}) which
 very similar to the Doppler shift of the energy $\epsilon_d=V_s
 k_z$ which would appear due to the superflow $V_s=\alpha_j k_Fq^2_\perp$
 along the vortex axis. However there is no real superflow
 in the situation that we consider. Still the order parameter symmetry
 in $w$ and $uvw$ vortices allows the appearance of a spontaneous
 superflow along the vortex axis \cite{VolovikRMP}
 and the effective Doppler shift term in the energy spectrum.

  In contrast to the $v$ vortex
  the spectra of $w$ and $uvw$ vortices (\ref{w-A+beta-spectrum},\ref{vw-A-phase-Spectrum})
  consist of to asymmetrical anomalous branches [see
  Fig.(\ref{Spectrum})d]. In particular it means that the slopes of two anomalous branches
  $d E/dQ$ can be different at the Fermi level $E=0$.
   As distinct from the case of singular $o$ and
   $u$ vortices the difference in slopes contains no small parameter
and therefore therefore can be of significant value.

Since the anomalous branches in spectra of nonsingular vortices
intersect the Fermi level at finite angular momenta $Q_{1,2}\neq
0$ there exists a large number of zero modes, i.e. the energy
states exactly at the Fermi level.  We have calculated the number
of zero modes for all three types of nonsingular vortices assuming
a model situation when only the additional A phase is present.
 In a qualitative agreement with the results of work
\cite{VolovikLocalizedFermions} the number of zero modes was shown
to be of the order $N_0\sim k_F\xi_v$, where $\xi_v$ is a size of
vortex core.

The significant modification of the spectra of bound fermions as
compared to the CdGM case should result in various ramifications
of the vortex dynamics which is governed by the kinetics of vortex
core quasiparticles \cite{KopninSalomaa}. With the help of the
analytical results for the spectra obtained in this paper it
should be possible to explore the dynamics of nonsingular vortices
in $^3$He-B.

  {\bf 5. Acknowledgements.}
 This work was supported, in part, by Russian Foundation for Basic Research,
 by Programs of RAS "Quantum Physics of Condensed Matter" and "Strongly
  correlated electrons in semiconductors, metals, superconductors and magnetic materials",
  and by "Dynasty" Foundation. It is my pleasure to thank G.E. Volovik and A.S.
  Mel'nikov for numerous stimulating discussions.

\end{document}